\documentclass[11pt]{article}
\usepackage[export]{adjustbox}
\usepackage{hyperref}

\usepackage{graphicx}
\usepackage[labelformat=simple]{subcaption}

\usepackage{algorithm, algorithmicx}
\usepackage{algpseudocode}
\usepackage{booktabs}
\usepackage{epstopdf}
\usepackage{fullpage}
\usepackage{cite}
\usepackage{balance}

\title{\textbf{\Large CRUM: Checkpoint-Restart Support for CUDA's Unified Memory}}
\date{}

\author{Rohan Garg \\
        Northeastern U. \\
        Boston, USA \\
        \url{rohgarg@ccs.neu.edu}
        \and
        Apoorve Mohan  \\
        Northeastern U. \\
        Boston, USA \\
        \url{apoorve@ccs.neu.edu}
        \and
        Michael Sullivan \\
        NVIDIA Corp. \\
        Santa Clara, USA \\
        \url{misullivan@nvidia.com}
        \and
        \\
        Gene Cooperman \\
        Northeastern U. \\
        Boston, USA \\
        \url{gene@ccs.neu.edu}}

\begin{document}
\maketitle
\begin{abstract}

Unified Virtual Memory (UVM) was recently introduced on recent NVIDIA GPUs.
Through software and hardware support, UVM provides a
coherent shared memory across the entire heterogeneous node, migrating data as
appropriate. The older CUDA programming style is akin to older large-memory
UNIX applications which used to directly load and unload memory segments. Newer
CUDA programs have started taking advantage of UVM for the same reasons of
superior programmability that UNIX applications long ago switched to assuming
the presence of virtual memory.  Therefore, checkpointing of UVM will become
increasingly important, especially as NVIDIA CUDA continues to gain wider
popularity: 87 of the top 500 supercomputers in the latest listings are
GPU-accelerated, with a current trend of ten additional GPU-based
supercomputers each year.

A new scalable checkpointing mechanism, CRUM (Checkpoint-Restart for
Unified Memory), is demonstrated for hybrid CUDA/MPI computations across
multiple computer nodes. CRUM supports a fast, forked checkpointing,
which mostly overlaps the CUDA computation with storage of the checkpoint
image in stable storage. The runtime overhead of using CRUM is 6\% on
average, and the time for forked checkpointing is seen to be a factor
of up to 40~times less than traditional, synchronous checkpointing.

\end{abstract}

\section{Introduction}

The advent of virtual memory automated the task of managing a program's memory
segments. Hence, for large, complex programs, the use of virtual memory becomes
{\em more efficient in practice}, since few programmers wish to spend
development time manually squeezing out the most efficient memory management.
In much the same way, NVIDIA has introduced {\em Unified Virtual Memory} (UVM)
into their recent GPUs.  CUDA UVM is analogous to the
virtual memory with hardware support found on traditional computers.

UVM is especially important for workloads with memory footprints that
are too large to entirely fit in device memory.
In this case, UVM allows the application to allocate its data within a UVM
region that is directly visible to a kernel running on the device.
A ``working set'' of memory is automatically paged into the device
as needed.  Furthermore, the use of a unified
virtual address space enables deployment of complex data structures for
GPU-based computation, with the same pointers being valid on the host as well
as on the GPU.

\begin{figure}[t!]
\centering
\includegraphics[scale=0.44]{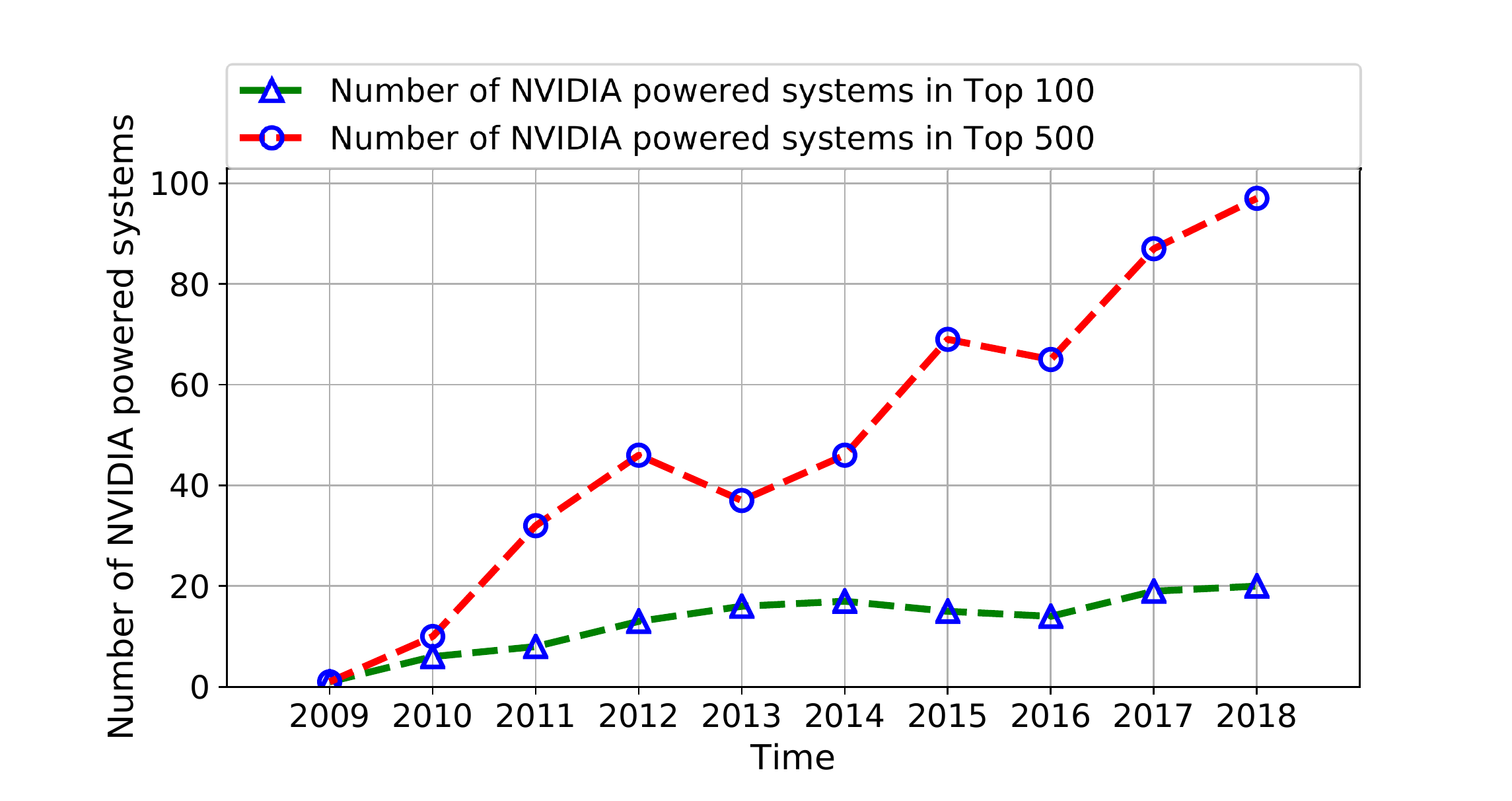}
  \caption{NVIDIA GPUs in Top 500 list.}
\label{fig:top}
  \vspace{-0.5cm}
\end{figure}

The use of GPUs continues to grow as seen in recent TOP-500
lists~\cite{top500-2017} (see Figure~\ref{fig:top}), and the advent of a
unified shared address space is expected to further lower the entry barrier
and widen the adoption of GPUs in HPC systems.

Unfortunately, GPUs have been shown to suffer from
a high rate of {\em Detected Unrecoverable Errors}
(DUEs)~\cite{haque2010hard,shi2011sustainable,debardeleben2014gpu,
tiwari2015reliability,tiwari2015understanding,sridharan2015memory}.  The mean
time between failures (MTBF) is expected to become much worse as the number of
compute nodes increases in the exascale generation.

Thus, efficient checkpointing for the UVM model is important for the future
exascale generation. Unfortunately, previous checkpointing
research~\cite{shi2009vcuda,gupta2009gvim,takizawa2009checuda,
gomez2010transparent,nukada2011nvcr,gtc2016crcuda}  assumes the older
(non-UVM) memory model.

A na{\"i}ve approach to support checkpoint-restart would be to: (a)
introspect and save the application process state (including the CUDA
user-space library) and the GPU device driver; and (b) restore the
process memory (including the CUDA user-space library) and restore the
GPU device driver state. Unfortunately, the CUDA user-space library,
which is checkpointed and restored as part of the process memory, is
non-reentrant. Thus, it cannot restore the GPU device driver state.

To address these challenges, this paper proposes a novel framework, CRUM
(Checkpoint-Restart for Unified Memory), which decouples the application process
state from the device driver state (see Section~\ref{sec:design}) by using a proxy process.
Thus, CRUM can transparently checkpoint the application without involving
any active driver state. (This could potentially allow a CUDA application to be
checkpointed on one version of CUDA and GPU hardware, and restarted on another
CUDA/GPU version.)

To optimize checkpointing of applications with large memory footprints, CRUM
uses fork-based, copy-on-write mechanism.
There are two phases. The first,
and relatively fast, phase is the transfer of data resident on the GPU
hardware to the application process through a proxy process.
In the second phase, the application process disconnects from
the proxy and forks a child process that writes the checkpoint data to stable storage.
Meanwhile, the application process re-connects to the proxy, which resumes
using the GPU for computation.

This work makes the following two novel contributions:

\begin{enumerate}
  \item An algorithm for {\em shadow page synchronization}
	(see Algorithm~\ref{algo:stateSyncAlgo}), which ensures
	the isolation of an application process
	from the GPU device, while allowing the UVM memory
	regions to be shared between the two; and
\item A {\em forked checkpointing} model for UVM memory that
	overlaps writing a checkpoint image to stable storage
	while the application continues.
	This was difficult previously
	due to the need to share memory between the GPU device and host (UVM),
	and simultaneously between parent and forked child process.
\end{enumerate}

Experimental results show that CRUM provides an effective and scalable
approach for checkpoint-restart of real-world, high-performance computing
workloads that take advantage of CUDA 8's UVM (Section~\ref{sec:evaluation}).
These hybrid CUDA/MPI applications include the DOE benchmarks HPGMG-FV and
HYPRE.  An average runtime overhead of 6\% was observed.  Further, CRUM's
fast, forked checkpointing reduces the time to checkpoint up to a factor of
40~times less than a traditional checkpoint that writes out process memory
to stable storage. CRUM is open source software that will be freely available.

Section~\ref{sec:motivation} presents the background and
motivation, including both the need for UVM support and the
need for greater GPU reliability as we approach the exascale generation.
Section~\ref{sec:design} describes the design of CRUM, while
Section~\ref{sec:evaluation} presents an experimental evaluation.
Section~\ref{sec:discussion} presents an analysis of the current
limitations of the current approach, and the potential impact on
future generations of NVIDIA GPUs.
Finally, Section~\ref{sec:relatedWork} describes the related work,
and Section~\ref{sec:conclusion} presents the conclusion.

\section{Background and Motivation}
\label{sec:motivation}

\subsection{History and Motivation for Unified Virtual Memory (UVM)}

Unified Virtual Memory (UVM) and its predecessor, Unified Virtual Addressing
(UVA), are major CUDA features that are incompatible with prior CUDA
checkpointing approaches. Yet, UVM is an important innovation for
future CUDA applications.

Through software and hardware support, UVM provides a
coherent shared memory across the entire heterogeneous
node~\cite{cuda_programming_2017,harris_2017}.  The use of UVM-managed memory
greatly simplifies data sharing and movement among multiple GPUs. This is
especially useful given that the most energy-efficient supercomputers place
multiple compute accelerators per node---for instance,
TSUBAME3.0~\cite{tiffany_trader_tsubame3.0}, Coral Summit~\cite{coral_summit},
and the NVIDIA SATURNV~\cite{roy_kim_nvidia_2016} supercomputer use 4, 6, and 8
GPUs per node, respectively. The features and progression of UVM are briefly
described below.

Historically, in CUDA 4 (2011), Fermi-class GPUs added support for Unified
Virtual Addressing
(UVA) with {\em zero-copy memory}.  UVA allows transparent zero-copy
accesses to memory across a
heterogeneous node using a partitioned address space. UVA never migrates data,
and so non-local memory accesses suffer from less bandwidth and longer latency.

\begin{figure}[t!]
\centering
\includegraphics{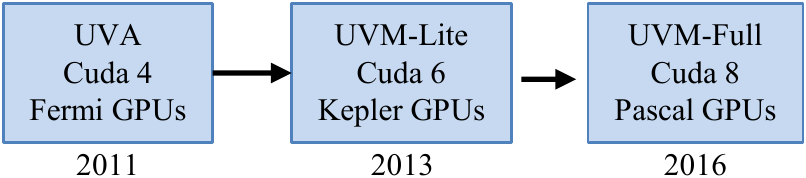}
\caption{The technology advancement of CUDA unified virtual memory\@.}
\label{fig:uvm_generations}
  \vspace{-0.5cm}
\end{figure}

To reduce the performance penalty of non-local zero-copy memory accesses,
first-generation Unified Virtual Memory (UVM-Lite) was introduced in CUDA 6
(2013) for Kepler-class GPUs~\cite{harris_2013}. UVM-Lite shares a single
memory space across a heterogeneous node, and it transparently migrates all
memory pages that are attached to the CUDA streams associated with each kernel.
This simplifies deep copies with pointer-based structures and it allows GPUs to
transparently migrate UVM-managed memory to the device, nearly achieving the
performance of CUDA programs using explicit memory management. Due to hardware
restrictions, however, UVM-Lite does not allow concurrent access to the same
memory from both CPU and GPU---host-side access is only allowed once all
GPU-side accesses to a CUDA stream have completed. Concurrent access to
UVM-managed memory from different GPUs is allowed, but data are never migrated
between devices and non-local memory is accessed in a zero-copy fashion.

Second-generation UVM (UVM-Full) was introduced in CUDA 8 (2016) for
Pascal-class GPUs~\cite{harris_2016}. It eliminates the concurrent-access
constraints of the prior UVM generation and adds support for system-wide atomic
memory operations, providing an unrestricted coherent shared memory across the
heterogeneous node. On-demand data migration is supported by UVM-Full across
all CPUs and GPUs in a node, with the placement of any piece of data being
determined by a variety of heuristics~\cite{harris_2017}.

Pascal-era UVM also
adds support for memory over-subscription, meaning that UVM-managed regions
that are larger than the GPU device memory can be accessed without explicit
data movement.  This is important for applications with large data.
In particular, it greatly simplifies the programming of large-memory jobs,
and avoids the need to explicitly marshal data to and from the
GPU~\cite{sakharnykh_amrproxy_2016}. For instance, GPU-capacity-exceeding deep
neural network training has been accomplished in the past through explicit data
movement~\cite{rhu_vdnn:_2016}, but it can also be performed with less
programmer effort by UVM over-subscription~\cite{sakharnykh_gtc_2017}.

\subsection{GPUs for Exascale: DUEs and GPU Reliability}
\label{sec:motivateReliability}

The advantages of using GPUs for high-performance computing have been realized and a steep rise in their use in large-scale HPC systems has been observed (see Figure~\ref{fig:top}). Eighty-seven (87) systems in the Top500 list were reported to be powered by NVIDIA GPUs in November 2017, as compared to one (1) in November 2009~\cite{top500-2017}. Thus, it is important that both hardware and the software stack (pertaining to the use of GPUs) should be highly available and reliable to maximize large-scale HPC systems productivity.

While this makes GPUs attractive for exascale computing, the high GPU detectable-uncorrectable error rate (as compared to CPUs) remains an issue.  Checkpointing
plays an important role in mediating this issue.
Various studies have been conducted for understanding the reliability aspects of using GPU's in large-scale HPC systems.
The studies suggest that the newer generation GPU's are more reliable, as are the large-scale HPC systems using them (i.e., the observed MTBF of systems using newer GPU's is much longer than their estimated MTBF)~\cite{haque2010hard,shi2011sustainable,debardeleben2014gpu,tiwari2015reliability,tiwari2015understanding,sridharan2015memory}.

However, one factor that motivates efficient checkpoint-restart on GPU accelerated systems is that GPU memory currently tends to have more DUEs
(Detected Unrecoverable Errors) per GB than CPU memory. Memory in CPU nodes is composed of narrow 4-bit or 8-bit wide DRAM devices that are grouped together into DIMMs, meaning certain ECC codes (often called chipkill ECC) can correct the data that comes from an entire DRAM device. In contrast, GPU memory is much wider (32-bit wide for GDDR5/GDDR5X and 128-bit for HBM2) such that chipkill-level protection is not possible without a prohibitively large memory access granularity; accordingly, current GPUs use single-bit correcting SEC-DED ECC for DRAM~\cite{nvidia_p100,oliveira2014gpgpus}. These lesser correction capabilities lead to a relative increase in detected errors. For example, a field study of the Blue Waters system~\cite{di2014lessons} found that the DUE rate per GB of Kepler-era GDDR5 was roughly 5~times that of the chipkill-protected CPU memory.

Given
the high rate of DUEs expected in the future exascale systems,
checkpoints will be more frequent, and so it is imperative
to design checkpointing mechanisms that can reduce the time that
applications spend in checkpointing.

\subsection{Checkpointing Large-memory CUDA-UVM Applications}

UVM acts as an enabler for easily developing large-memory CUDA applications.
UVM enables a GPU to transparently access host CPU and remote GPU memory, and
hence solves the problem of otherwise manually managing data transfers.
All of the host CPU's memory is available, on-demand, by the GPU
device. Conversely, all of the UVM memory on the GPU device is available to
the CPU.

In this situation, the CUDA application may use much more memory than
is present on the device.
The capacity of GPU memory is currently from 16 to 32~GB for a high-end GPU,
while CPU memory often ranges from 128 to 256~GB.  In the past, this
forced GPU application developer to choose between:  scaling out to many nodes
and GPUs (hence incurring communication overhead); or manually managing the
data transfers on a single GPU.  Later, UVM made possible a third choice:
transparently transferring data on a single GPU
via UVM.  However, the ease of developing such large-memory CUDA-UVM
applications now places a larger burden on transparent checkpointing to
support this large-memory overhead.

\section{CRUM: Design and Implementation}
\label{sec:design}

To address the challenges described in Section~\ref{sec:motivation}, this
paper proposes CRUM, a novel framework that provides a checkpointing-based
fault-tolerance mechanism. CRUM enables transparent, system-level
checkpointing for CUDA and CUDA UVM applications.

\begin{figure}[ht]
  \centering
        \begin{subfigure}[t]{0.48\textwidth}
                \centering
                \includegraphics[scale=0.15]{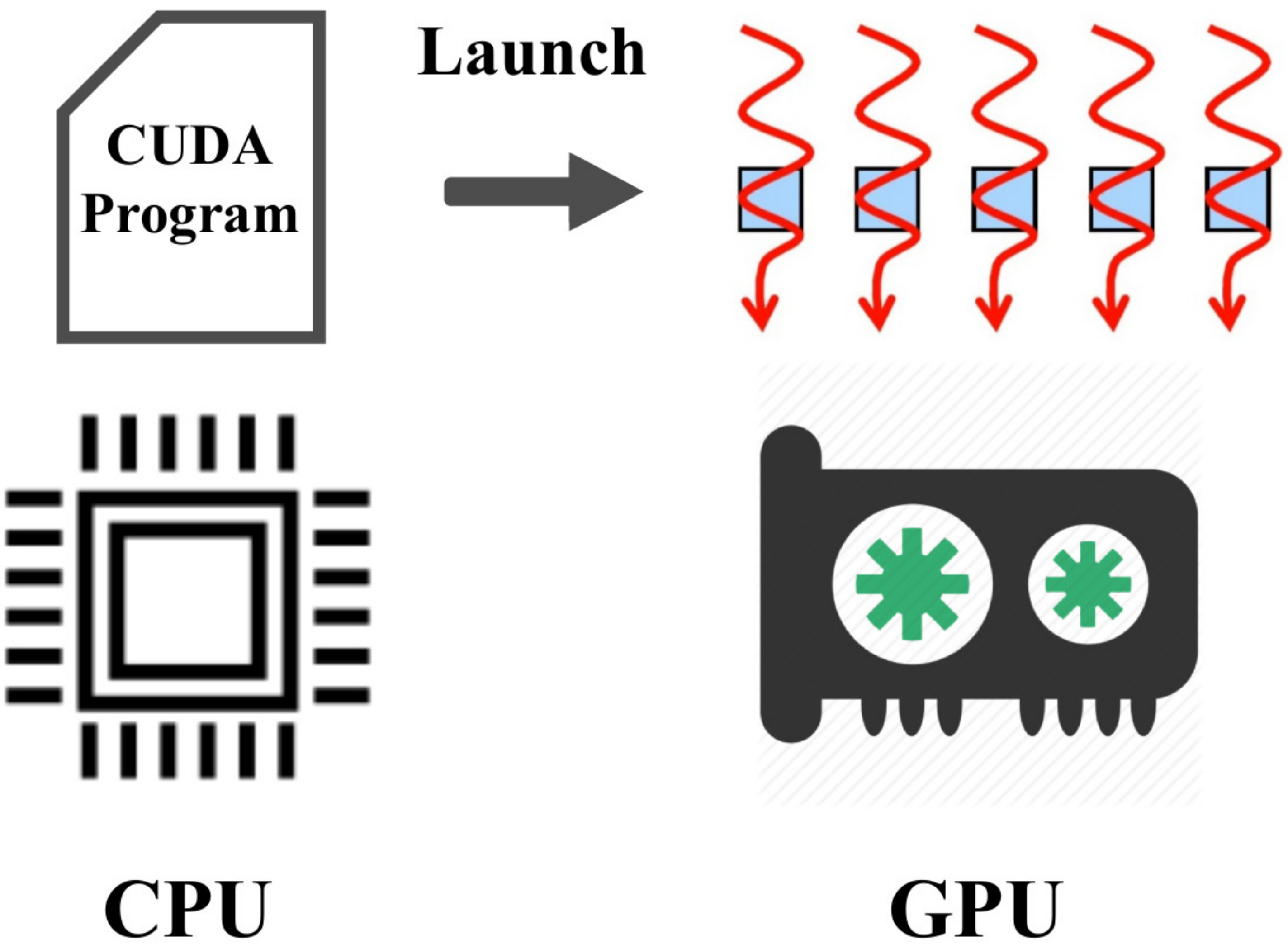}
                \caption{CUDA Original} \label{fig:cuda-orig}
                \vspace{6mm}
        \end{subfigure}
        \begin{subfigure}[t]{0.48\textwidth}
                \centering
                \includegraphics[scale=0.15]{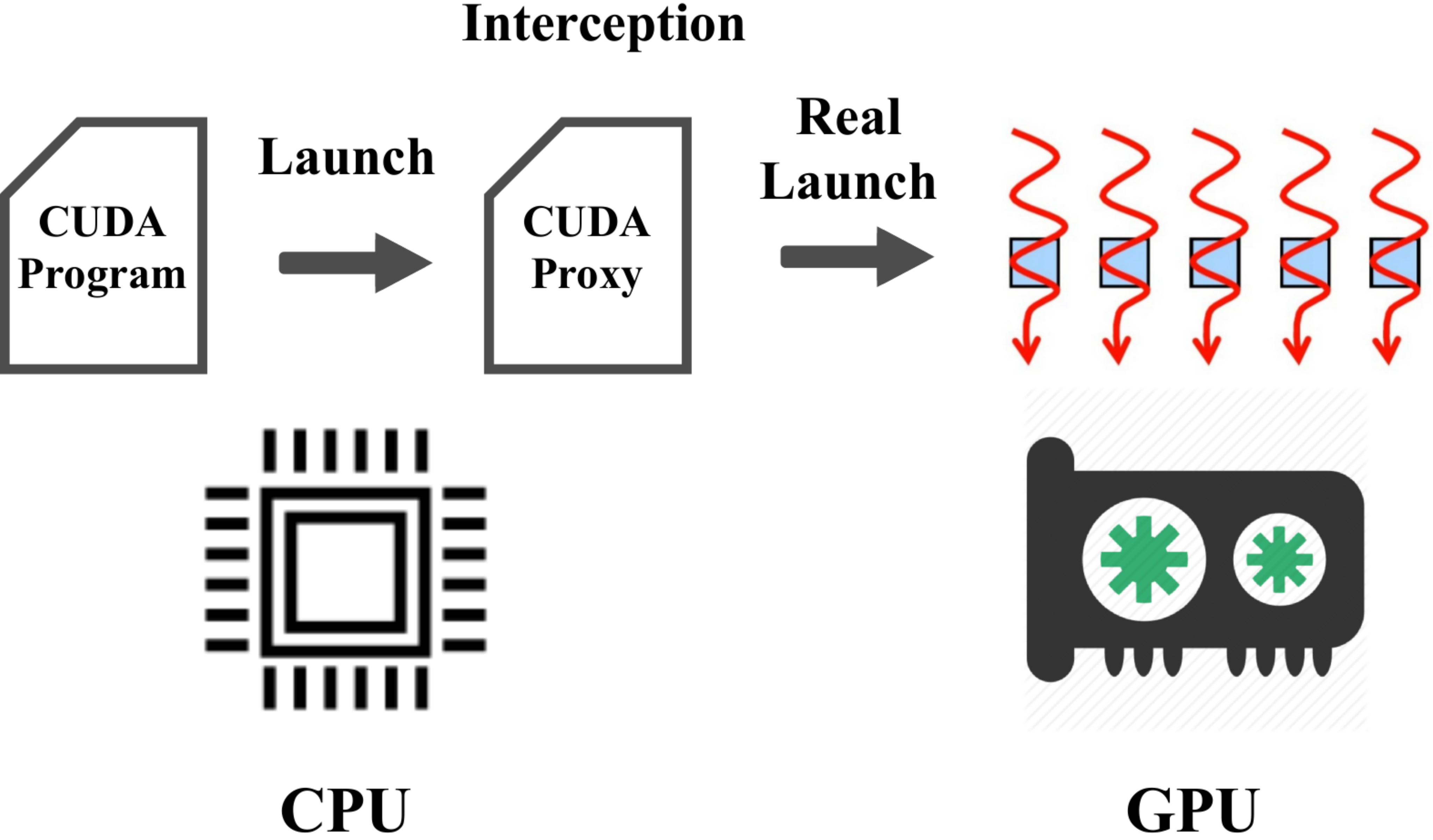}
                \caption{CUDA Proxy} \label{fig:cuda-proxy}
        \end{subfigure}
        \caption{High-level architecture of CRUM}
        \label{fig:cuda-orig-proxy}
\end{figure}

Figure~\ref{fig:cuda-orig-proxy} shows a high-level schematic of CRUM's
architecture. Note especially the organization into two processes:  a
CUDA program (the user's application), and a CUDA proxy (the only process
that uses the CUDA library to communicate with the GPU).  The flow of
control is: (i)~to interpose on CUDA library calls made by the application
process; (ii)~to forward the requests to the proxy process; (iii)~which
then executes the calls via its CUDA library and GPU, on behalf of the
application; and (iv)~finally returns the results back to the application.

In this section, we present the key subsystems in the design
of CRUM.  The first research challenge is the propagation of UVM
memory pages (already shared between GPU hardware and proxy process)
to make them visible to the application process.
Section~\ref{sec:shadow-UVM} describes a shadow page scheme
(summarized in Algorithm~\ref{algo:stateSyncAlgo})
for this purpose.  The second research challenge is to extend
this scheme to overlap checkpointing and computation for the
sake of fast, forked checkpoint and future exascale needs.
This is discussed in Section~\ref{sec:forkedCkpt}.
Finally, the implementation details of integrating CRUM
with proxy processes is discussed in~\ref{sec:proxyCkptRestart}.

\subsection{Post-CUDA~4:  The Need for a Proxy Process}
\label{sec:proxyMotivation}

Ideally, a single-process approach toward checkpointing seems simpler.
But this approach for CUDA became non-viable with CUDA~4 and beyond, when
NVIDIA implemented unified virtual addressing with zero-copy, an antecedent
of unified memory~\cite{sakharnykh_gtc_2017}).  At that point,
it was no longer possible to re-initialize the CUDA library at the time
of restart.  We assume that this is due to the lack of clear semantics
about what it means to re-initialize a CUDA library that still retains
pointers to unified memory regions on host and device.  One must choose
either to free the host memory (thus sabotaging any CUDA application that
retains a pointer to the unified memory region), or else to leave
the host memory region intact (thus sabotaging any application assumptions
about unification of host and device memory).  Note that a fresh restart
will restore all host memory, but any unification of host with
device memory has already been lost.

The core issue is that the CUDA
unified memory model was developed for standard CUDA applications --- and
naturally did not include extensions for transparent checkpointing.
An alternative workaround would have been, at restart time, to overwrite
the text and data memory segments of any CUDA libraries with a fresh,
uninitialized CUDA library (matching a freshly booted GPU), and then to
call {\tt cudaInit()}.  Unfortunately, the CUDA library/driver appeared to have
additional state, which made this workaround infeasible.

\subsection{Shadow Pages for the Support of UVM}
\label{sec:shadow-UVM}

Recall the use of a proxy process, as seen in
Figure~\ref{fig:cuda-proxy}.  The core research challenge in this
architecture is that UVM dictates that pages are transparently
shared between the GPU hardware and the proxy process, but these
shared UVM pages are not visible to the application process.

The zero-copy memory of CUDA~4 implies that there are no CUDA calls
on which to interpose.  In direct-mapped memory, the device may
read or write to the host mapped pinned memory of the proxy process
at any time.  But the separate application process remains unaware
of modifications to memory in the proxy process.  Thus, an approach
using CUDA proxies is unable to support the newer and potentially
more efficient zero-copy memory for UVA. To overcome this situation,
a new, transparent checkpointing approach for CUDA's zero-copy
memory is proposed, in which proxy and application reflect a single
application with two ``personalities''.

The CUDA application process and the CUDA proxy process invoke the
same application binary but execute two different state machines.
The application process goes through three different states: CUDA
call, read from device-mapped UVM memory, write to device-mapped
UVM memory. Note that the state transitions are not dictated by
the CRUM framework, but rather by the application logic.  On the
other hand, the CUDA proxy process is simply a passive listener
for requests from the application process and executes the CUDA calls
and the memory reads and writes as dictated by the application.

Based on these observations, we introduce the concept of ``shadow
UVM pages''. For every CUDA UVM allocation request by the application,
CRUM creates a corresponding shadow UVM region in the context of
the application process. At the same time, the CUDA proxy process
requests for a ``real'' UVM region from the device driver. The two
processes, the {\em application} and the {\em proxy}, see two different
views of the memory and data at any given point.

Since there are no API calls to interpose on, this opens up the
requirement for tracking the changes to the application process's
memory in order to keep the two sets of pages in sync. CRUM relies
on the use of user-space page-fault tracking to accomplish this. There
are currently two available mechanisms for page-fault tracking in
Linux: \texttt{userfaultfd}; and segfault handler and page protection
bits. While there are certain performance benefits with the use of
\texttt{userfaultfd}, the current work uses segfault handler and
page protection bits to allow for evaluation on clusters employing
older Linux kernels.

The algorithm for synchronizing the data on shadow and real UVM
pages is described in Algorithm~\ref{algo:stateSyncAlgo}.

\algdef{SE}[EVENT]{Event}{EndEvent}[1]{\textbf{upon event}\ #1\ \algorithmicdo}{\algorithmicend\ \textbf{event}}%
\algtext*{EndEvent}

\begin{algorithm}
\begin{algorithmic}
  \caption{\label{algo:stateSyncAlgo} Shadow page synchronization algorithm}

  \Event{Page Fault}
   \If {addr $\in$ AllShadowPages}
     \If {isReadFault()}
       \State ReadDataFromRealPage()
     \Else
       \State MarkPageAsDirty()
     \EndIf
   \EndIf
  \EndEvent

  \Event{CUDA call}
    \If {hasDirtyPages}
      \State SendDataToRealPages()
      \State ClearDirtyPages()
    \EndIf
  \EndEvent

  \Event{CUDA Create UVM region}
    \State uvmAddr $\gets$ CreateUvmRegionOnProxy()
    \State reg $\gets$ CreateShadowPage(uvmAddr)
    \State AllShadowPages $\gets$ AllShadowPages $\cup$ reg
  \EndEvent

\end{algorithmic}
\end{algorithm}

When an application process requests for a new UVM region, a new
shadow UVM region is created in the process's memory (using the
\texttt{mmap} system call). The shadow UVM region is given read-write
permissions initially, and all the pages in the regions are marked
as ``dirty''.

When the application makes a CUDA call where the device could
potentially read or modify the UVM data (for example, a CUDA kernel
launch), the data from dirty pages is ``flushed'' to the real UVM
pages on the proxy process, the dirty flag is cleared for the UVM
region, and the read-write permissions are removed (using the
\texttt{mprotect} system call).

This allows CRUM to interpose on either a read or write to unified
memory. Standard Linux code for segfault handlers allows CRUM to
detect an attempt to read or to write, and to distinguish the two
cases. In the case of a read, PROT\_READ permission is set for all
of the memory in the application process corresponding to unified memory.
In the case of a write, PROT\_WRITE permission is set
for all of the memory in the application process corresponding to unified
memory. (See Section~\ref{sec:linuxPagePerms} for further discussion.)

At a later time, when the application process tries to read the
results of the GPU computation back from the shadow UVM regions, a
read page fault is triggered; the permissions of the shadow UVM region
are changed to read-only, and the results are read in from the
corresponding real UVM region on the proxy.

\subsubsection{Page permissions on Linux}
\label{sec:linuxPagePerms}

Note that write to shadow UVM memory region requires PROT\_WRITE permission.
Unfortunately, on Linux, PROT\_WRITE permission implies PROT\_READ
permission also.  Linux does not support {\em write-only} permission,
but rather {\em read-write} permission instead.

This has consequences for the three-state algorithm to support
unified memory in CRUM.  We make the assumption here that most
applications will cycle through the three states in order (possibly
omitting the read-only or write-only phase).  Hence, a typical cycle
would be invoked: CUDA-call/read-unified-memory/write-unified-memory.

In fact, CRUM also supports overlapped execution of a CUDA call
with reading and writing unified memory.  The essential assumption
is that read access must precede write access and a read-write cycle
cannot be followed with a second read unless there is an intervening
CUDA kernel.  Normal CUDA calls such as \texttt{cudaMemcpy} are allowed at
all times.

As discussed earlier, unfortunately, Linux's write-only permission
for memory actually grants read-write permission.  It is for this reason
that a transition from write-unified-memory directly to read-unified-memory
cannot be detected efficiently.  Possible solutions are discussed at the
end of this section.

This assumption has been found to hold in each of the real-world
applications that we have found for testing CRUM with unified memory.
Nevertheless, it is important to also build in a (possibly slower)
{\em verified execution mode} that will test an
application to see if it violates this assumed cycle of
CUDA-call/read-unified-memory/write-unified-memory.

There is more than one way to implement a verified execution mode.

One of the difficulties is that a Linux segfault handler
does not allow us to reset the page permission to allow only the
pending read or write, and then reset the permission back to
PROT\_NONE.  Linux's user-space page fault handling, \texttt{userfaultfd}, introduced
with Linux 4.3, can fix this, but that introduces other technical
difficulties.  (For example, it was only with Linux 4.11 that this
was extended partially to support fork and shared memory.)  Another
alternative is to parse the pending read or write (load or store assembly
instruction), temporarily allow read-write permission to the desired
memory page, and then use the parsed information to read or write the
data between register and memory, and finally to restore the previous
memory permission. This might be more efficient than user-space page
faulting since it might have fewer transitions to a kernel system call.

Linux kernel modification to support write-only permissions for UVM
shadow pages is another possibility.

\subsection{Fast, Forked Checkpoints}
\label{sec:forkedCkpt}

UVM enables CUDA applications to use all of the host and GPU device memory
transparently. This can make checkpointing, which is dominated by the time
to write to the disk, prohibitively expensive. So while one could employ
copy-on-write-based asynchronous checkpointing, UVM memory is incompatible
with shared memory and fork on Linux.

Fortunately, CRUM's proxy-based architecture can be used to address this challenge.
Note that the device state and the UVM memory regions are not directly a part
of the application process's context, but rather they are associated with the
proxy process. This frees up the application process to use forked checkpointing
for copy-on-write-based associated checkpointing for the application process.

Forked checkpointing allows CRUM to invoke a minimal checkpointing delay in order
to ``drain the GPU device'' of its data, after which, a child process of each
MPI process can write to stable storage. This allows the system to overlap
the CUDA computation with storage of the checkpoint image in stable storage.

\subsection{Checkpoint-Restart Methodology and Integration with Proxies}
\label{sec:proxyCkptRestart}

Finally, for completeness, we discuss how CRUM integrates proxy
concepts into the CUDA implementation requirements.  Proxies have
also been used by previous authors (see Section~\ref{sec:relatedWork}-d).

At checkpoint time, CRUM suspends the user application threads, and
``drains'' the GPU kernel queue. It issues a device synchronize call
(\texttt{cudaDeviceSynchronize}) to ensure that the kernels have finished
execution and the memory state is consistent. Then, for all the active
CUDA-MALLOC
and CUDA-UVM memory regions, data is read in from the GPU to the host. The
data is first transferred from the GPU into the proxy process's memory,
and then from the memory of the proxy process into the memory of the user
application process.  The user application process then disconnects from the
proxy process. This ensures that the problem reduces to the trivial problem
of checkpointing a single-process application. Finally, the state of the
process is saved to a checkpoint image file on stable storage.

At the time of restart, CRUM starts a new process and recreates the user
application threads. Then, the memory of the new process gets replaced by
the saved state from the checkpoint image file. CRUM, then, starts a new
proxy process, which starts a new GPU context. It recreates the active
CUDA-MALLOC and CUDA-UVM memory regions by replaying the allocation calls.
CUDA streams and events are similarly handled.
(See Section~\ref{sec:discussion} for further discussion.)
Finally, CRUM transfers
the data into the actual CUDA and CUDA-UVM regions through the proxy process
and resumes the application threads.

\section{Experimental Evaluation}
\label{sec:evaluation}

The goal of this section is to present a detailed analysis of the performance
of CRUM. In particular, this section answers the following questions:

\textbf{Q1} {\itshape What's the overhead of running a CUDA (or a CUDA UVM) application under CRUM?}

\textbf{Q2} {\itshape Does CRUM provide the ability to checkpoint CUDA (and CUDA UVM)
applications?}

\textbf{Q3} {\itshape Can CRUM improve a CUDA UVM based application's throughput by
reducing the checkpointing overhead?}

\textbf{Q4} {\itshape Is the approach scalable?}

\subsection{Setup}

To answer the above questions, we first briefly describe our
experimental setup and methodology.

\subsubsection{Hardware}

The experiments were run on a local cluster with 4 nodes. Each node
is equipped with 4 NVIDIA PCIe-attached Tesla P100 GPU devices, each with 16~GB of RAM. 
The host machine is running a 16-core Intel Xeon E5-2698 v3 (2.30~GHz) processor with 256~GB of RAM. Each node runs CentOS-7.3 with Linux
kernel version 3.10.

\subsubsection{Software}

Each GPU runs NVIDIA CUDA version 8.0.44 with driver 396.26.
Experiments use DMTCP~\cite{ansel2009dmtcp} version~3.0. We developed a
CRUM-specific DMTCP plugin~\cite{arya2016design} for checkpoint-restart of
NVIDIA CUDA UVM applications.

The DMTCP CRUM plugin (referred to as the CRUM plugin from here
onwards) interposes on the CUDA calls made by the application. The
interposition code is generated in a semi-automated way, where a
user specifies the prototype of a CUDA function, and whether the
call needs to be logged. This not only allows us to cover the
extensive CUDA API, but also allows for ease of maintainability and
for future CUDA extensions.

The plugin forwards the requests, over a SysV shared memory region, to a proxy
process running on the same node. The forwarded request is then executed by the
proxy process, which then returns the results back to the application. To
improve the performance, we use well-studied concepts from pipelining of
requests,
to allow the application to send requests without blocking. Blocking requests,
such as, \texttt{cudaDeviceSynchronize}, result in a pipeline flush. For data
transfers (both for UVM shadow page data and for {\tt cudaMalloc} data) we use
Linux's
Cross Memory Attach (CMA) to allow for data transfers using a single copy
operation.

\subsubsection{Application Benchmarks}
\label{sec:benchmarks}

We use Rodinia~3.1~\cite{che_rodinia_2009} benchmarks for evaluating CRUM for
CUDA applications. Note that the Rodinia benchmarks do {\em not} use UVM, and
can be run even with CUDA 2.x.  They are included here to show comparability of
the new approach with the older work from 2011 and earlier using CUDA
2.x~\cite{takizawa2009checuda,nukada2011nvcr,takizawa2011checl}.

We note that CheCUDA~\cite{takizawa2009checuda} does not work for
modern CUDA (i.e., CUDA version 4 and above) because it relies on a
single-process checkpoint-restart approach. CheCL~\cite{takizawa2011checl}
only supports OpenCL and does not work with CUDA. We tried compiling the
CRCUDA~\cite{gtc2016crcuda} version available online~\cite{crcudaSource},
but it failed to compile with CUDA version 8. It didn't work for the
benchmarks used in our experiments, after applying our compilation fixes.

To evaluate CRUM using UVM-managed memory allocation, we run a GPU-accelerated
build of two DOE benchmarks: a high-performance geometric multigrid proxy
application (HPGMG-FV~\cite{hpgmg}), and a test application using a
 production linear system solver library (HYPRE~\cite{hypre}).  For the
HYPRE library, we run the test driver for an unstructured matrix interface
using
the AMG-PCG solver. For HPGMG-FV, we evaluate two versions: the standard
HPGMG-FV benchmark with one grid (the {\em master} branch, as described
in~\cite{sakharnykh_hpgmg_2016}), and an AMR proxy modification with multiple
AMR levels (the {\em amr\_proxy} branch, as described in~\cite{sakharnykh_amrproxy_2016}).

We focus on HPGMG-FV and HYPRE because they are scientific applications and
libraries with potential importance in future exascale
computing~\cite{kothe_douglas_exascale_2016}, and
they have publicly available ports to UVM-enabled multi-GPU CUDA. HPGMG-FV has
also been used as a benchmark for ranking the speeds of the top
supercomputers~\cite{hpgmg_benchmarking}.

To evaluate the relative performance of HPGMG-FV runs, we quote its throughput
in
degrees-of-freedom per second --- the same metric used to rank supercomputer
speeds~\cite{hpgmg_benchmarking}.  Thus, larger numbers indicate higher
performance.
To evaluate the relative performance of HYPRE
runs, we measure the wall clock time taken by each program execution.

\subsection{Runtime Overhead}

While the ability to checkpoint is important for improving the
throughput of an application on a system with frequent failures, a
checkpointing system that imposes excessive runtime overhead can render the
framework ineffective, and in the worst case, reduce the throughput.
We, therefore, benchmark and
analyze the sources of runtime overhead.  For these experiments, no
checkpoint or restart was invoked during the run of the application.

The results demonstrate that CRUM is able to run the CUDA application with a
worst case overhead of 12\%, and a 6\% overhead on average. We note that this
is
a prototype implementation and a production system could incorporate many
optimizations to further reduce the overhead.

\begin{table}[ht]
\caption{\label{tbl:rodiniaConfig} Runtime parameters for Rodinia applications.}
\centering
  \begin{tabular}{|l|c|}
  \hline
  Application & Configuration Parameter \\
  \hline
    LUD       & ``-s 2048 -v'' \\
  \hline
    Hotspot3D & ``512 8 1000 power\_512x8 temp\_512x8'' \\
  \hline
    Gaussian  & ``-s 8192'' \\
  \hline
    LavaMD    & ``-boxes1d 40'' \\
  \hline
\end{tabular}
\end{table}

\begin{figure*}[t!]
  \begin{subfigure}[b]{0.32\textwidth}
    \centering
    \includegraphics[height=120pt, width=\textwidth]{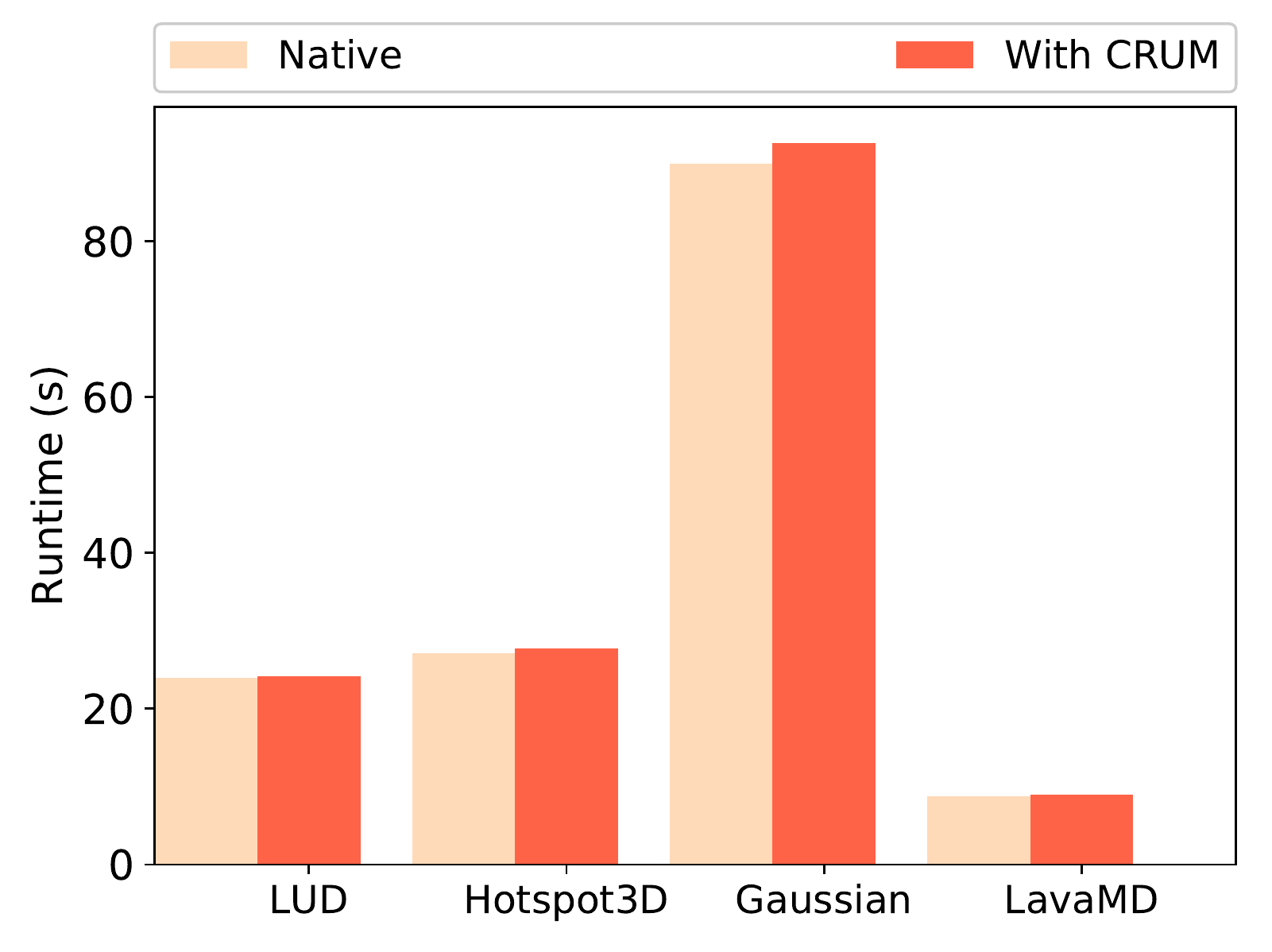}
    \caption{Rodinia. \label{fig:rodiniaRuntime}}
  \end{subfigure}
  \begin{subfigure}[b]{0.32\textwidth}
    \centering
    \includegraphics[height=120pt, width=\textwidth]{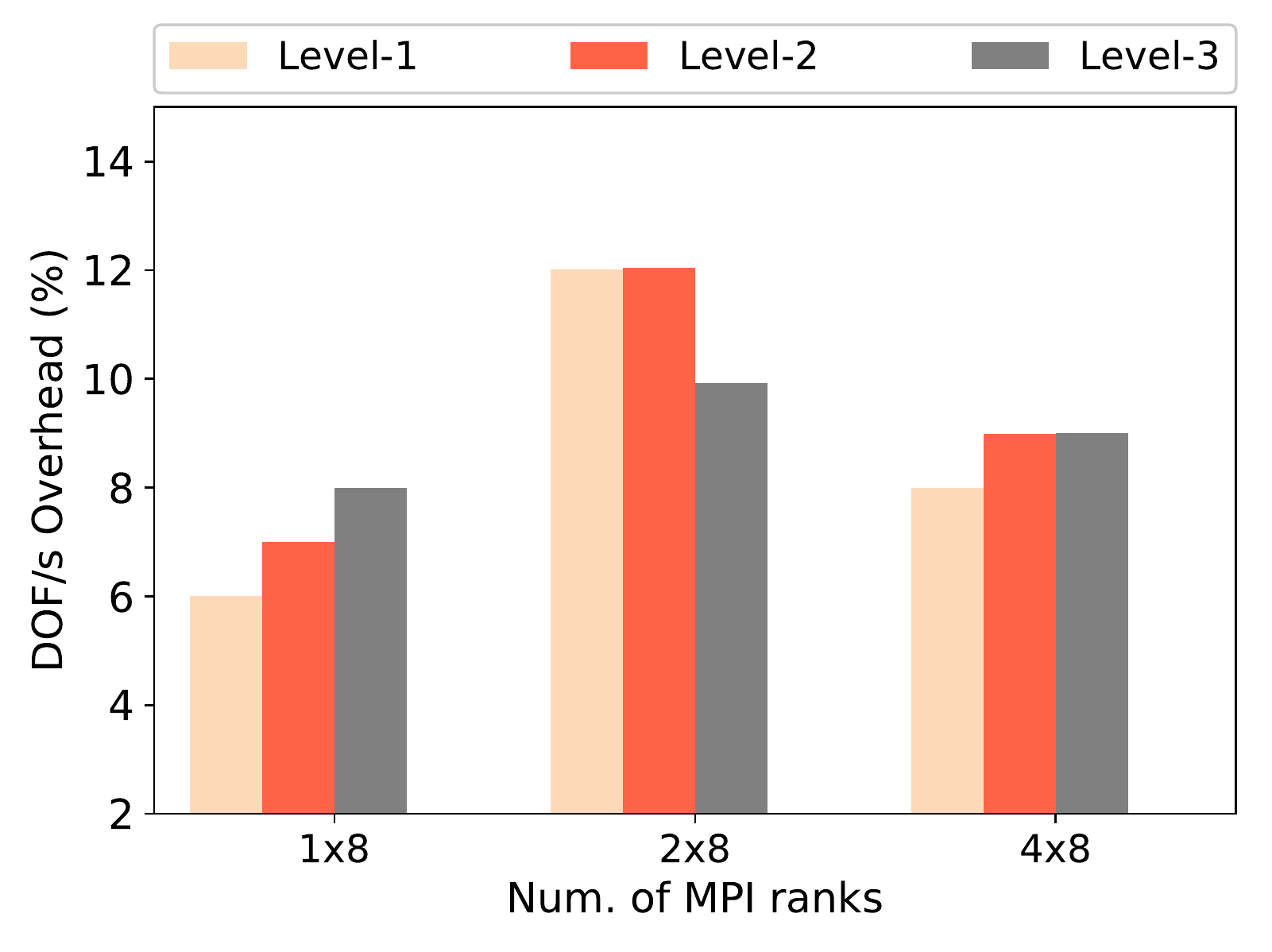}
  \caption{HPGMG-FV. \label{fig:hpgmgThroughput}}
  \end{subfigure}
  \begin{subfigure}[b]{0.32\textwidth}
    \centering
    \includegraphics[height=120pt, width=\textwidth]{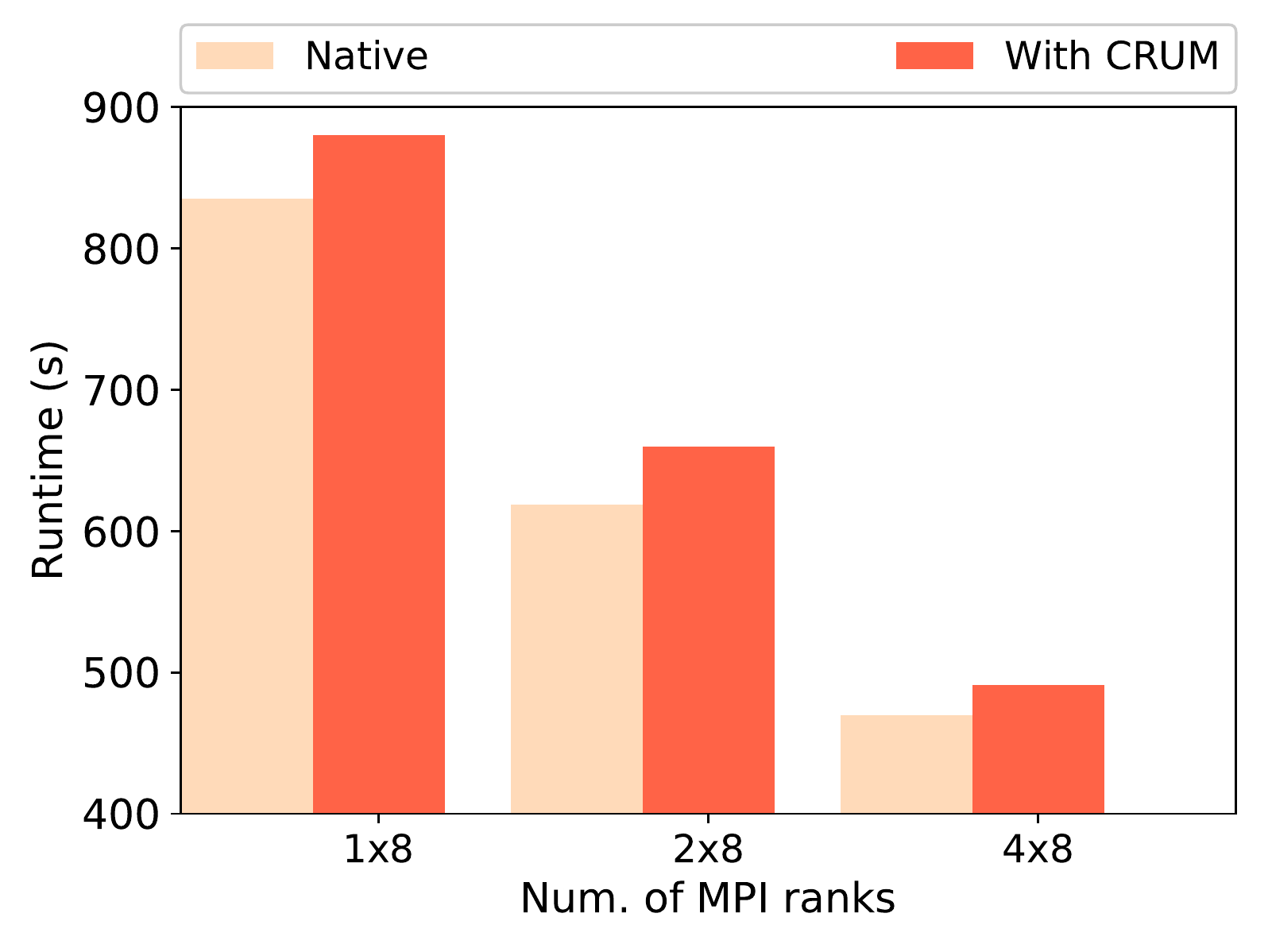}
    \caption{HYPRE. \label{fig:hypreRuntime}}
  \end{subfigure}
  \caption{Runtime overheads for different benchmarks under CRUM.}
\end{figure*}

Figure~\ref{fig:rodiniaRuntime} shows the runtimes for four applications (LUD, Hotspot3D,
Gaussian, and LavaMD) from the Rodinia benchmark suite with and without CRUM.
The applications mostly use the CUDA API's from CUDA 2.x: {\tt cudaMalloc}, {\tt
cudaMemcpy}, and {\tt cudaLaunch}. Table~\ref{tbl:rodiniaConfig} shows the
configuration parameters used for the experiments. We observe that the runtime
overhead varies from 1\% (for LUD) to 3\% (in the case of LavaMD). The runtime
overhead is dominated by the cost of data transfers from the application process
to the proxy process. In a different experiment, using Unix domain sockets
for data transfer, we observed overheads varying from 1.5\% to 16.5\%. The use of
CMA reduces the overhead significantly.

Figure~\ref{fig:hpgmgThroughput} shows the runtime results for the HPGMG-FV benchmark with
increasing number of nodes and MPI ranks. As noted in
Section~\ref{sec:benchmarks}, we use the HPGMG-FV throughput metric  DOF/s as a proxy for performance. We note that the DOF/s reported by
the application running under CRUM are less than the native numbers by 6\% to 12\%.
We present a more in-depth analysis below.

In our experiments, we observed that a single MPI rank of the HPGMG-FV benchmark
runs about 9 million CUDA kernels during its runtime of 3 minutes. This implies
that each CUDA kernel runs for approximately 20 microseconds on average. Note that
the cost of executing a {\tt cudaLaunch} call itself can be up to 5 microseconds. The
program allocates many CUDA UVM regions, sets up the data, and runs a series of
kernels to operate on the data. Each MPI rank then exchanges the results with
its neighbors. While the size of the UVM regions vary from 12~KB to 128~KB, the
frequent reads and writes the application process, stresses the CRUM framework
in two dimensions: (a) frequent interrupts and data transfer; and (b) frequent
context switches and the need to synchronize with proxy process (because of the
many CUDA calls that need to be executed).

While the use of CMA (cross-memory attach) reduces the cost of data transfers,
interestingly, we
observed a lot of variability in the cost of a single CMA operation for the same
data transfer size. The cost of a single page transfer varies from 1 microsecond
to 1 millisecond, a difference of three orders of magnitude. We attribute this
to two sources: (a) O/S jitter; (b) the pre-fetching algorithm employed by the
UVM driver. In many cases, reading a UVM page is slowed down because of a
previous read on a large UVM region, spanning several pages, because the driver
gets busy pre-fetching the data for the large UVM region.

To address the second source of overhead, we optimized the CRUM implementation
to: (a) use a lock-free, inter-process synchronization mechanism over
shared-memory; and (b) pipeline non-blocking CUDA calls from the application. A
CUDA call, such has {\tt cudaLaunch}, {\tt cudaMemsetAsync}, is pipelined and
the application is allowed to move ahead in its execution, while the proxy
finishes servicing the request. At a synchronization point, like {\tt cudaDeviceSynchronize},
the application must wait for a pipeline flush, i.e., for the pending requests
to be completed.

Figure~\ref{fig:hypreRuntime} shows the runtimes for the HYPRE benchmark for a different
number of MPI ranks running on a varying number of nodes. The benchmark observes
up to 6.6\% overhead when running under CRUM compared to native execution.

The HYPRE benchmark presents different checkpointing challenges than HPGMG-FV.  While the HYPRE benchmark
invokes only about 100 CUDA kernels per second (10 milliseconds on average per
kernel) during its execution, it uses many large UVM regions (up to 900~MB).
Thus, the overhead is dominated by the cost of data transfers between the application
process and the proxy.

In addition to CMA, CRUM employs a simple heuristic to help reduce the data
transfer overhead.  For small shadow UVM regions, it reads in all of the data
from the real UVM pages on the proxy.  However, for a read fault on a large
shadow UVM region, it starts off by only reading the data for just one page
containing the faulting address. On subsequent read faults on the same region,
while in the read phase (see Section~\ref{sec:design}), we exponentially
increase (by powers of 2) the number of pages read in from the real UVM region
on the proxy. This heuristic relies on the spatial and temporal locality of
accesses. While there will be pathological cases where an application does
``seemingly'' random reads from different UVM regions, we have found this
assumption to be valid in the two applications we tested.

\subsection{Checkpointing CUDA Applications:  Rodinia and MPI}

Next, we evaluate the ability of CRUM to provide fault tolerance for CUDA and
CUDA UVM applications using checkpoint-restart.

\begin{figure*}[t!]
  \begin{subfigure}[b]{0.32\textwidth}
    \centering
    \includegraphics[height=120pt, width=\textwidth]{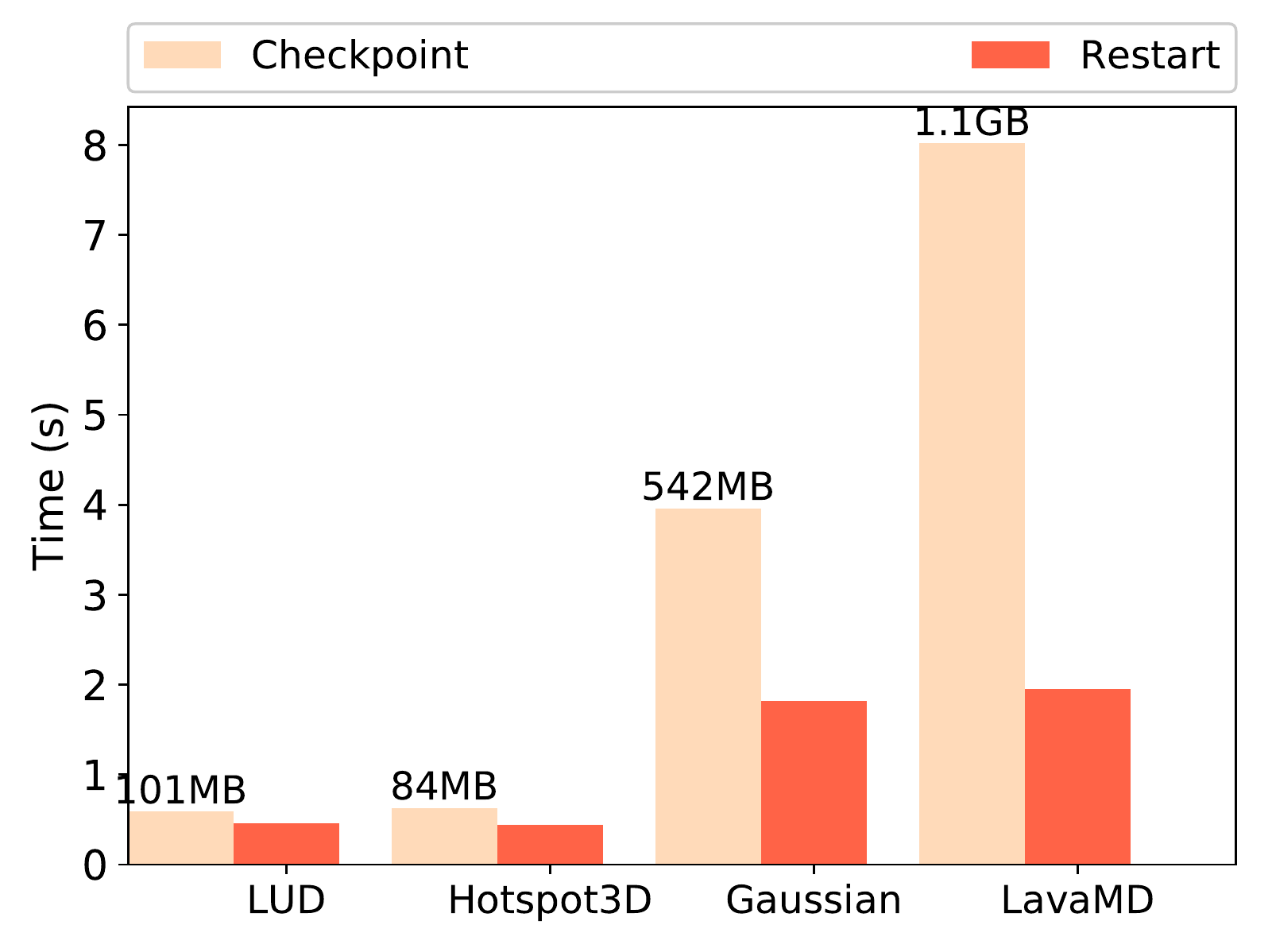}
    \caption{Rodinia. \label{fig:rodiniaCkpttime}}
  \end{subfigure}
  \begin{subfigure}[b]{0.32\textwidth}
    \centering
    \includegraphics[height=120pt, width=\textwidth]{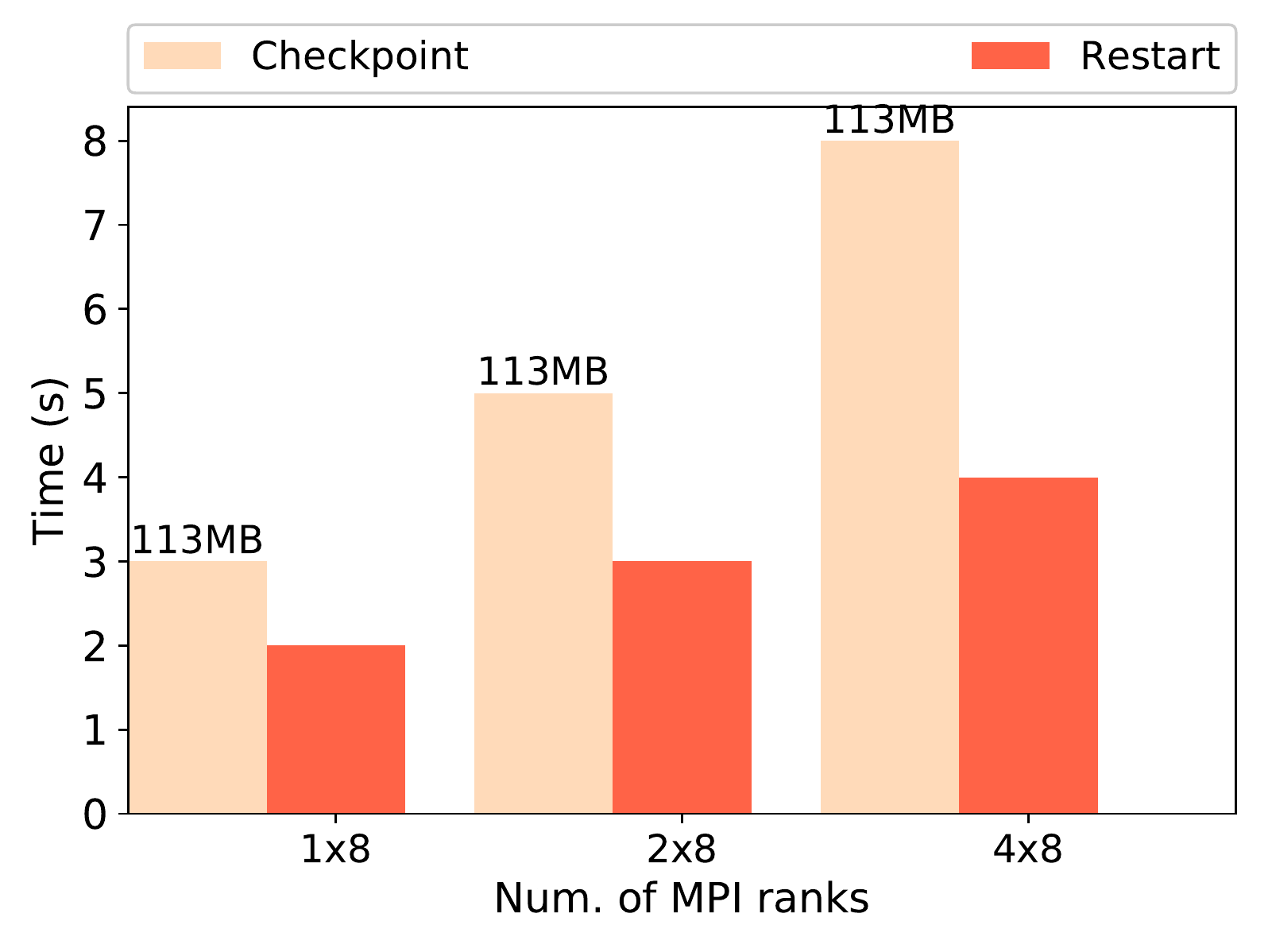}
    \caption{HPGMG-FV. \label{fig:hpgmgCkpttime}}
  \end{subfigure}
  \begin{subfigure}[b]{0.32\textwidth}
    \centering
    \includegraphics[height=120pt, width=\textwidth]{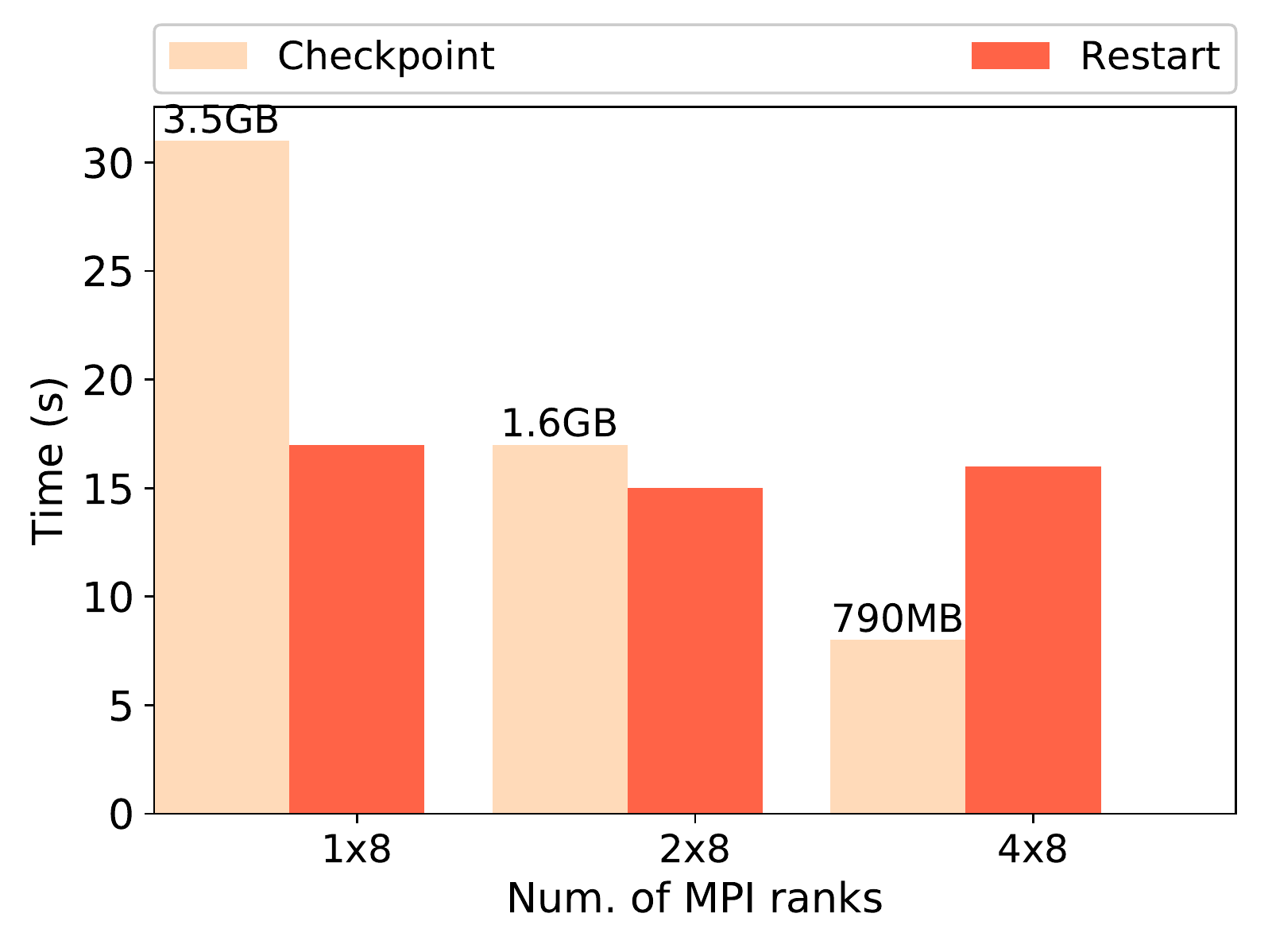}
    \caption{HYPRE. \label{fig:hypreCkpttime}}
  \end{subfigure}
  \caption{Checkpoint-restart times and checkpoint image sizes for different benchmarks under CRUM.}
\end{figure*}

Figure~\ref{fig:rodiniaCkpttime} shows the checkpoint times, restart
times, and the checkpoint image sizes for the four applications from
the Rodinia benchmark suite.
The checkpointing overhead is dependent on the time to transfer the data from
the device memory to the host memory, then transferring it from the proxy process
to the application process using CMA, and then finally writing to the disk.
We observe that the time to write dominates the checkpointing time.

Figure~\ref{fig:hpgmgCkpttime} shows the checkpoint times, restart times, and the
checkpoint image sizes for HPGMG. The results are shown with increasing number
of MPI ranks (and the number the nodes). We observe that as the total amount of
checkpointing data increases from 904~MB ($8 \times 113$~MB) to 3.6~GB ($32
\times 113$~MB), the checkpoint time increases from 3 seconds to 8 seconds. We
attribute the small checkpoint times to the buffer cache on Linux. We observed
that forcing the files to be synced (by using an explicit call to {\tt fsync}
increased the checkpoint times by up to 3~times.

The results for HYPRE are shown in Figure~\ref{fig:hypreCkpttime}. The application divides
a fixed amount of data (approx. 28~GB in total) equally among its ranks. So, we
observe that the checkpoint image size reduces by almost half every time we
double the number of ranks.  This helps improve the checkpoint cost especially
with smaller process sizes, as the Linux buffer caches the writes, and the
checkpoint times reduce from 31 seconds (for 8 ranks on 1 node) to 8 seconds
(for 32 ranks over 4 nodes). 

\subsection{Reducing the Checkpointing Overhead:
	  A Synthetic Benchmark for a Single GPU}

To showcase the benefits of using CRUM to reduce checkpointing overhead
for CUDA UVM applications,
we develop a CUDA UVM synthetic benchmark. The synthetic benchmark allocates two vectors of
$2^{32}$ 4-byte floating point numbers (32~GB in total) and computes the dot
product of the two vectors. The floating point numbers are generated at random.
Note that the total memory requirements are double of what is available on the
GPU device (16~GB). However, UVM allows an application to use more than the
available memory on the GPU device. The host memory, in this case, acts as ``swap
storage'' for the device and the pages are migrated to the device or to the host
on demand.

\begin{table}[ht]
  \caption{\label{tbl:synthCkpting} Checkpoint times using different strategies for the synthetic benchmark.}
\centering
  \begin{tabular}{|l|r|c|c|}
  \hline
    Strategy & Ckpt Time & Ckpt Size & Data Migration\\
             &           &           &  Time   \\
  \hline
    Na{\"i}ve  & 45~s  & 33~GB (100\% random) & 4~s \\
  \hline
    Gzip  & 1296~s & 29~GB (100\% random) & 4~s \\
  \hline
    Parallel gzip  & 86~s & 29~GB (100\% random) & 4~s \\
  \hline
    LZ4  & 62~s & 33~GB (100\% random) & 4~s \\
  \hline
    Forked Ckpting & 4.1~s & 32~GB (100\% random) & 4~s \\
  \hline
    Gzip  & 749~s  & 15~GB (50\% random) & 4~s \\
  \hline
    Parallel gzip  & 56~s  & 15~GB (50\% random) & 4~s \\
  \hline
    LZ4  & 45~s  & 17~GB (50\% random) & 4~s \\
  \hline
\end{tabular}
\end{table}

Table~\ref{tbl:synthCkpting} shows the checkpoint times for three different
cases: (a)~using a na{\"i}ve checkpointing approach; (b)~using three
different compression schemes, Gzip, Parallel Gzip, and LZ4, before
writing to the disk; and (c)~using CRUM's forked checkpointing approach.
The first two approaches, na{\"i}ve and compression, use CRUM's CUDA UVM
checkpointing framework.  The third approach adds the forked checkpointing
optimization to the base CUDA UVM checkpointing framework. The three compression
schemes use Gzip's lowest compression level (\texttt{-1}~flag).
While parallel Gzip
uses the same compression algorithm as Gzip, it launches as many threads as
the number of cores on a node to compress input data.

We observe that the forked checkpointing approach outperforms the other two
approaches by up to three orders of magnitude.  Since the program uses random
floating point numbers, compression is ineffective at reducing the size of
the checkpointing data (Table~\ref{tbl:synthCkpting}). We note that the time
taken by the
compression algorithm is also correlated with the randomness of data. As an
experiment, we introduced redundancy in the two input vectors to improve the
``compressibility''. Of the $2^{32}$ floating point elements in a vector, only
half ($2^{16}$) of the elements were generated randomly and the rest were
assigned the same floating point number. This improves the compression time and
reduces the checkpoint time to 749 seconds and the checkpoint image size is
reduced to 15~GB by using the Gzip-based strategy.

Note that parallel Gzip may not be a practical option in many HPC scenarios,
where an application often uses one MPI rank per core on a node. On the other
hand, LZ4 provides a computationally fast compression algorithm at the cost
of a lower compression ratio.

\subsection{Reducing the Checkpoint Overhead:  Real-world MPI Applications}

Finally, we present the results from using CRUM with the forked checkpointing
optimization for the real-world CUDA UVM application benchmarks. The results
reported
here correspond to the largest scale of 4 CPU nodes, with 16 GPU devices,
running 8~MPI ranks per node (32~processes in total).

\begin{table}[ht]
  \caption{\label{tbl:realWorldCkpting} Checkpoint times using different strategies for real-world CUDA UVM applications. The numbers
  reported corresponds to running 32 MPI ranks over 4 nodes. The checkpoint size reported is for each MPI rank. The checkpoint times are normalized to the time for the na{\"i}ve checkpointing approach (1x).}
\centering
  \begin{tabular}{|l|l|c|r|}
  \hline
    App. & Strategy & Ckpt Time & Ckpt Size  \\
  \hline
    HPGMG-FV & Gzip  & 0.78x & 14~MB  \\
  \hline
    HPGMG-FV & Parallel gzip  & 0.60x & 14~MB  \\
  \hline
    HPGMG-FV & LZ4  & 0.30x & 16~MB  \\
  \hline
    HPGMG-FV & Forked ckpting & 0.025x  & 113~MB  \\
  \hline
    HYPRE & Gzip  & 2x  & 176~MB  \\
  \hline
    HYPRE & Parallel gzip  & 1x  & 176~MB  \\
  \hline
    HYPRE & LZ4   & 1x  & 296~MB  \\
  \hline
    HYPRE & Forked ckpting  & 0.032x  & 868~MB  \\
  \hline
\end{tabular}
\end{table}

Table~\ref{tbl:realWorldCkpting} shows the results for checkpointing time (and checkpoint
image sizes) normalized to the checkpointing time using the na{\"i}ve
checkpointing approach (as shown in Figures~\ref{fig:hpgmgCkpttime} and~\ref{fig:hypreCkpttime}). The results are
shown for HPGMG-FV and HYPRE.

We observe trends similar to the synthetic benchmark case. While in the na{\"i}ve
checkpointing approach, the checkpointing overhead is dominated by the cost of
I/O, i.e., writing the data to the disk, under forked checkpointing, the
overhead is dominated by the cost of in-memory data transfers: from the GPU to
the proxy process, and from the proxy process's address space to the
application
process's address space. Further, the cost of quiescing the application
process,
quiescing the network (for MPI), and ``draining'' and saving the in-flight
network
messages is 0.01\% of the total cost.

However, unlike the synthetic benchmark, using in-memory compression to reduce
the size of data for writing is better in this case for both HPGMG and HYPRE.
This indicates that the compression algorithm is able to efficiently reduce the
size of the data, which helps lower the I/O overhead. Note that this is still
worse than using forked checkpointing by an order of magnitude.

\section{Discussion}
\label{sec:discussion}

{\bf Driver support for restart:} In order to restart a computation, CRUM must
re-allocate memory in the same locations as during the original
execution---otherwise the correctness of pointer-based code cannot be
guaranteed during re-execution. The current CRUM prototype relies on
deterministic CUDA memory allocation, which we verify to work with the CUDA
driver libraries via experimentation (for both explicit device memory and
UVM-managed memory allocation). The assumption of deterministic memory
re-allocation is shared by previous GPU checkpointing
efforts~\cite{nukada2011nvcr}.

{\bf Memory Overhead:} In a CUDA program with large
data resident on the host, the memory overhead due to an additional
proxy process could be a concern.  In the special case of asynchronous
checkpointing, the overhead could be even higher, although copy-on-write
does prevent it from going too high. This could be ameliorated by future
support for shared memory UVM pages between application and a proxy
running CUDA.

{\bf Advanced CUDA language features:} Dynamic parallelism allows CUDA kernels
to recurse and launch nested work; it is supported by CRUM without change.
Device-side memory allocation allows kernel code to allocate and de-allocate
memory. It is partially supported by CRUM, with one important distinction---no
live device-side allocations are allowed at a checkpoint time.  Thus,
device-side memory allocations are to be freed before the system is considered
quiesced and ready for a checkpoint. We do not anticipate this constraint to be
particularly difficult to satisfy, since device-side mallocs tend to be used to
store temporary thread-local state within a single kernel, whereas host-side
CUDA memory allocation (which is supported by CRUM without restriction) is more often used for persistent storage.

\textbf{Using \texttt{mprotect}:} Currently, in a Linux kernel, {\tt PROT\_WRITE}
protection for a memory region implies read-write memory permission rather than
write-only memory permission.  Because of this, some compromises were needed in
the implementation.  This work has demonstrated the practical advantages of a
write-only memory permission for ordinary Linux virtual memory.  It is hoped
that in the future, the kernel developers at NVIDIA will be encouraged to
support write-only memory permission for this purpose.

Another issue with an mprotect-based approach is that when kernel-space
code page faults on a read/write protected page, it returns an error code
to the user, {\tt EFAULT}, rather than a segfault. This forces the implementation
to be extended to handle such failures; the implementation cannot rely solely
on a segfault handler~\cite{plank1994libckpt,ruscio2007dejavu,bugnion2013lightweight,vogt2015lightweight}.

{\bf Other APIs and Languages:} This work provides checkpoint-restart
capabilities for programs written in C/C++ with the CUDA runtime library.  In
our experience, the CRUM prototype should support the majority of
GPU-accelerated HPC workloads; however, there are other APIs to that may be
valuable for some users. Given the current framework of code auto-generation for
CRUM, we believe that it will be straightforward to extend the implementation to support
other APIs, such as OpenACC.  The ability of CRUM to support UVM-managed memory
would be especially useful for OpenACC programs, as PGI's OpenACC compiler
provides native and transparent support for high-performance UVM-managed
programs, making UVM-accelerated OpenACC programs a low-design-effort route to
performant GPU acceleration~\cite{sakharnykh_openacc_2015}.

{\bf Future Versions of CUDA:} Just as prior checkpointing methods for GPUs
were unable to cope with versions of CUDA since CUDA~4 (released in 2011), it
is likely that CRUM will need to be updated to support language features after
CUDA 8. One such development is Heterogeneous Memory Management
(HMM)~\cite{hubbard_hmm_2017}, which is a kernel feature introduced in Linux
4.14 that removes the need for explicit cudaMallocManaged calls (or use of the
\_\_managed\_\_ keyword) to denote UVM-managed data. Rather, with HMM the GPU
is able to access any program state, including the entire stack and heap.
Because the current CRUM prototype relies on wrapping cudaMallocManaged calls,
it will need to be redesigned to support HMM.

\section{Related Work}
\label{sec:relatedWork}

\paragraph{Use of proxy process} Zandy et al.~\cite{zandy1999process}
demonstrated the use of a ``shadow'' process for checkpointing currently
running application processes that were not originally linked with a
checkpointing library. This allows the application process to continue to
access its kernel resources, such as open files, via RPC calls with the
shadow process.

Kharbutli et al.~\cite{kharbutli2006comprehensively} use a proxy process
for isolation of heap accesses by a process and for containment of attacks
to the heap.

\paragraph{GPU virtualization}
A large number of previous HPC studies have focused on virtualizing
the access to the
GPU~\cite{lagar2007vmm,shi2009vcuda,gupta2009gvim,takizawa2009checuda,giunta2010gpgpu,takizawa2011checl,nukada2011nvcr,gtc2016crcuda}.
Here we describe some of those studies, with an emphasis on
the use for GPU checkpointing and GPU-as-a-Service in the cloud and HPC
environments.

Lagar-Cavilla et al.~\cite{lagar2007vmm}, Shi et al.~\cite{shi2009vcuda},
Gupta et al.~\cite{gupta2009gvim}, and Giunta et al.~\cite{giunta2010gpgpu}
focus on providing access to the GPU for processes running in a
virtual machine (VM), as an alternative to PCI pass-through. The
access is provided by forwarding GPU calls to a proxy process that
runs outside the VM and has direct access to the GPU.

\paragraph{GPU-as-a-Service}
Two other efforts, DS-CUDA~\cite{oikawa2012dscuda} and rCUDA~\cite{duato2010rcuda},
have focused on providing access to a remote GPU for the purposes of
GPU-as-a-Service~\cite{reano2015local,reano2015performance,varghese2015acceleration,silla2016remote,reano2017intra,reano2017enhancing,prades2017turning}.
They also rely on a proxy process. Using the proxy process is similar to the
one described in this work; however, the focus is on efficient
remote access by using the InfiniBand's RDMA API. To the best of our
knowledge, none of the previous studies solve the problem of efficient
checkpointing of modern CUDA applications that use UVM. We note that
the optimizations described in these works can be used in conjunction
with CRUM for providing efficient access to remote GPUs.

\paragraph{GPU Checkpointing}
\label{sec:gpuCkpt}
Early work on virtualizing or checkpointing
GPUs was based on CUDA~2.2 and
earlier~\cite{shi2009vcuda,gupta2009gvim,takizawa2009checuda,
gomez2010transparent,nukada2011nvcr}.  Those approaches stopped working
with CUDA~4 (introduced in 2011), which introduced Unified Virtual Addressing
(UVA).  Presumably, it is the introduction of UVA that made it impossible
to re-initialize CUDA~4.

In 2016, CRCUDA~\cite{gtc2016crcuda}, employed a proxy-based approach,
similar to the 2011 approach of CheCL~\cite{takizawa2011checl}
that targeted OpenCL~\cite{stone2010opencl} (as opposed to CUDA) for GPUs.
OpenCL does not support unified memory, and so
CheCL and CRCUDA do not support NVIDIA's
unified memory~\cite{sakharnykh_gtc_2017} targeted here.

VOCL-FT~\cite{pena2015voclft} aims to provide resilience against
soft errors.  VOCL-FT leverages the OpenCL programming model to
reduce the amount of data movement: both to/from the device from/to
the host, and to/from the disk. This allows them to do fast checkpointing
and recovery.

HiAL-Ckpt~\cite{xu2010hialckpt}, HeteroCheckpoint~\cite{kannan2014heterocheckpoint}, and
cudaCR~\cite{pourghassemi2017cudaCR} use application-specific
approaches for providing GPU checkpointing.

None of the approaches described above work for CUDA UVM. CRUM focuses on
providing efficient runtime and checkpointing support for CUDA and CUDA-UVM
based programs. We note that the techniques described in above approaches are
complementary to CRUM and can be used to further optimize the runtime and
checkpointing overheads.

\section{Conclusion}
\label{sec:conclusion}

This paper introduced CRUM, a novel framework for checkpoint-restart for
CUDA's unified memory. The framework employs a proxy-based architecture
along with a novel shadow page synchronization mechanism to efficiently
run and checkpoint CUDA UVM applications. Furthermore, the architecture
enables fast, copy-on-write-based, asynchronous checkpointing for large-memory
CUDA UVM applications. Evaluation results with a prototype implementation
show that average runtime overhead imposed is less than 6\%, while improving
the checkpointing overhead by up to 40 times.

\section*{Acknowledgment}

We thank Nikolay Sakharnykh from NVIDIA for sharing
his knowledge of UVM programming, and the NVIDIA PSG Cluster for
compute time. We also thank Kapil Arya for insightful discussions and
his feedback on an earlier draft of the paper. And we wish to thank
Onesphore Ndayishimiye for his early prototype of an auto-generator for
communication between application and proxy process.

This work was partially supported by NSF Grants ACI-1440788 and OAC-1740218,
and by
Grant 2014-345 from a ``Chaire d'attractivit\'e'' de l'IDEX, Universit\'e
F{\'e}d{\'e}rale Toulouse Midi-Pyr{\'e}n{\'e}es.

\end{document}